\documentclass[12pt]{JHEP} 
\usepackage{epsfig}
\newcommand\fverb{\setbox\pippobox=\hbox\bgroup\verb}
\newcommand\fverbdo{\egroup\medskip\noindent%
			\fbox{\unhbox\pippobox}\ }
\newcommand\fverbit{\egroup\item[\fbox{\unhbox\pippobox}]}
\newbox\pippobox

\def\del{\partial}

\def\gap#1{\vspace{#1 ex}}

\def\pht{{\tilde \phi}}
\def\D{D}

\def\u{u}

\def\myitem#1{\gap2\noindent\underbar{#1}\gap2}

\def\be{\begin{equation}}
\def\ee{\end{equation}}
\def\ba{\begin{array}{l}}
\def\ea{\end{array}}
\def\bea{\begin{eqnarray}}
\def\eea{\end{eqnarray}}
\def\eq#1{(\ref{#1})}

\def\nn{\nonumber\\}

\def\eps{{\epsilon}}
\def\ket#1{| #1 \rangle}

\def\ads{$AdS_5 \times S^5$}

\title{Fermions from Half-BPS Supergravity}
\author{Gautam Mandal$^*$  \\
{Perimeter Institute of Theoretical Physics,\\
31 Caroline Street North\\ 
Ontario, Canada N2L 2Y5\\ 
($^*$On leave from Tata Institute of Fundamental Research)
}
\\~~\\
\email{gmandal@perimeterinstitute.ca, mandal@theory.tifr.res.in
}}

\preprint{\hepth{0502104}\\
\\
}

\abstract{We discuss collective coordinate quantization of the
half-BPS geometries of Lin, Lunin and Maldacena (hep-th/0409174). The
LLM geometries are parameterized by a single function $u$ on a
plane. We treat this function as a collective coordinate. We arrive at
the collective coordinate action as well as path integral measure by
considering D3 branes in an arbitrary LLM geometry. The resulting
functional integral is shown, using known methods (hep-th/9309028), to
be the classical limit of a functional integral for free fermions in a
harmonic oscillator. The function $u$ gets identified with the
classical limit of the Wigner phase space distribution of the fermion
theory which satisfies $u * u = u$. The calculation shows how
configuration space of supergravity becomes a phase space (hence
noncommutative) in the half-BPS sector. Our method sheds
new light on counting supersymmetric configurations in supergravity.}


\begin{document}

\section{Introduction} 

Recently it has been shown in \cite{LLM} that the half-BPS IIB
supergravity solutions, which are asymptotically $AdS_5 \times S^5$
and preserve an $O(4) \times O(4)$ symmetry of the asymptotic isometry
group, are in one-to-one correspondence with semiclassical
configurations of free fermions in a harmonic oscillator potential.
This result is yet another striking evidence of the AdS/CFT
correspondence \cite{Aharony:1999ti}, since the free fermions are
equivalent to \cite{Berenstein:2004kk} the half-BPS sector of the
super Yang-Mills theory. Related work can be found in
\cite{Corley:2001zk,Berenstein:2004kk,deMelloKoch:2004ws,Buchel:2004mc,Filev:2004yv,Bena:2004td,Suryanarayana:2004ig,Gauntlett:2004hs,Liu:2004ru,Martelli:2004xq,Burrington:2004hf,Chong:2004ce,Liu:2004hy,Sheikh-Jabbari:2005mf,Ebrahim:2005uz}.

The correspondence between the supergravity configurations and
semiclassical fermion configurations is based on a proposed
identification between a supergravity mode $u(x_1, x_2)$ with the
phase space density $u(q,p)$ of the free fermions, where $x_1,x_2$ are
two of the coordinates of the LLM geometry and $q,p$ are coordinates
of the phase space of the free fermions.  The present work began with
the questions (a) how two coordinates of space time can become phase
space (noncommutative) coordinates and (b) whether one can derive the
noncommutative dynamics directly from supergravity.

The plan of the paper is as follows. In Section 2, we mention a few
results of \cite{LLM} to identify the moduli space of half-BPS
vacua. The moduli space is parameterized by a single function
$u(x_1,x_2)$ (discussed in the previous paragraph) subject to two
constraints. In Section 3 we quantize the half-BPS configurations by
identifying $u$ as the collective coordinate. We provide a
parameterization of the generic function $u$ subject to the
constraints and identify them with D3 branes coupled to LLM
geometries. The collective coordinate actions are then calculated by
computing the D3 brane actions.  We use the formalism of phase space
path integrals to demonstrate how the phase space dimensions get
reduced by half under the BPS constraint and the configuration space
itself becomes a phase space. In Section 4 we collect the results and
rewrite the action as well as the measure in terms of the
$u$-variable. In Section 5 we identify the $u$-functional integral
with the classical limit of a functional integral describing free
fermions in a harmonic oscillator. In Section 6 we discuss a first
principles approach to derivation of the $u$-functional integral using
the general formalism of collective coordinates in the presence of BPS
constraints, using Kirillov's symplectic form. Section 7 contains a
summary and some open questions.  In Appendix A we present some
details concerning identification of the collective coordinate action
of Section 4 with the D3-brane actions of Section 3. Appendix B makes
a qualitative identification between gravitons and collective
excitations in the form of ripples.

Transformation of configuration space into a phase space under BPS
conditions has been considered in \cite{DJM} in the case of a giant
graviton probe in \ads. Supertubes have been discussed in somewhat
related contexts in \cite{Palmer:2004gu,Bak:2004kz}.  Rather appealing
similarities with parts of the present work can be found in
discussions on topological string/field theories
\cite{Dijk-Vafa,Aganagic:2003db,Vafa-talk}. Related ideas have also
appeared in the context of quantum hall systems in
\cite{Susskind:2001fb,Sheikh-Jabbari:2001au}.

\section{The moduli space of 1/2-BPS Supergravity }

As shown in \cite{LLM}, the half-BPS geometries (with $O(4) \times
O(4)$ symmetry) are characterized by a single function $z_0(x_1,x_2)
\equiv z(x_1,x_2,y=0)$ (see eqs. (2.5)-(2.15) of \cite{LLM}).  The
moduli space of these solutions is the space of $z_0$'s, subject to the
following regularity and topological constraints.

\myitem{The regularity constraint}

The constraint of regularity on the half-BPS geometries 
implies that $z_0$ can only be either $1/2$ or $-1/2$, that is
\footnote{$\chi_R(x)$ denotes the
characteristic function  of a region $R \subset {\bf R}^2$: 
\be
\chi_R(x) = 1 \ \hbox{if} \ x \in R, = 0\  \hbox{otherwise}.
\ee}
\be
z_0(x_1,x_2) = - \frac12\sum_i \chi_{R_i} + \frac12 \sum_j \chi_{\tilde R_j} 
\label{z-char}
\ee
where the $x_1,x_2$ plane is tessellated by the regions $R_i,\tilde R_j$,
with $z_0=-1/2$ in $R_i$ and  $z_0=1/2$ in the $\tilde R_j$. 

It is useful to define the function 
\be
\u(x_1, x_2) \equiv 1/2 - z_0(x_1, x_2)
\ee
The regularity constraint now reads $ \u(x_1, x_2) = 0 \ \hbox{or} \ 1
$, equivalently\footnote{$0 < u < 1$ gives rise to singular
solutions; e.g.\cite{Caldarelli:2004mz} identified the superstar solution 
\cite{Myers:2001aq,Cvetic:1999xp,Behrndt:1998jd} with
$0<u<1$. \cite{Caldarelli:2004mz} also showed in specific examples  
that the geometries with $u>1$ develop closed timelike curves.}
\be
\left(\u(x_1, x_2)\right)^2 = \u(x_1, x_2)
\label{quad-const}
\ee
The equation \eq{z-char} becomes 
\be
\u = \sum_i \chi_{R_i}(x_1,x_2)
\label{u-char}
\ee
where $R_i$ now denote regions with $u=1$.

\myitem{The topological constraint}

The topological constraint becomes \cite{LLM} 
\bea
\int_{R_i}\frac{dx_1dx_2}{2\pi \hbar} 
&&= N_i
\nn 
\int_{-\infty}^\infty \frac{dx_1dx_2}{2\pi \hbar} 
\u 
&&= \sum_i  N_i = N
\label{int-const}
\eea
where
\be
\hbar = 2\pi g_s \alpha'^2
\label{hbar-llm}
\ee
The condition that the geometries are asymptotically $AdS_5 \times
S^5$  implies that $R= \cup R_i$ is a
bounded region of the $x_1,x_2$ plane.

The functions $u(x_1,x_2)$ subject to the constraint
equations \eq{quad-const} and \eq{int-const} characterize
all regular half-BPS solutions of the system with $O(4) \times
O(4)$ symmetry and \ads\ asymptotics. 

\section{Quantization of half-BPS vacua}

We will treat the function $u$ as the collective coordinate of the
space of half-BPS configurations (with $O(4) \times O(4)$
symmetry). The space of $u$'s can be discussed in terms of orbits of a
specific $u_0$ under the action of the group of area-preserving
diffeomorphisms in two dimensions (see Section \ref{remarks} for this
description). Alternatively, $u$ can be parameterized as in
\eq{u-char}. By choosing generic enough regions $R_i$, we can describe
all functions $u$ subject to the constraints. This is the description
we will use in this and the following two sections to quantize the
space of $u$'s.

Let us choose
the regions as follows (see figure \ref{checkerboard.fig}):
\begin{figure}[ht]
\vspace{0.5cm}
\hspace{-0.5cm}
\centerline{
    \epsfxsize=8.5cm
   \epsfysize=8cm
   \epsffile{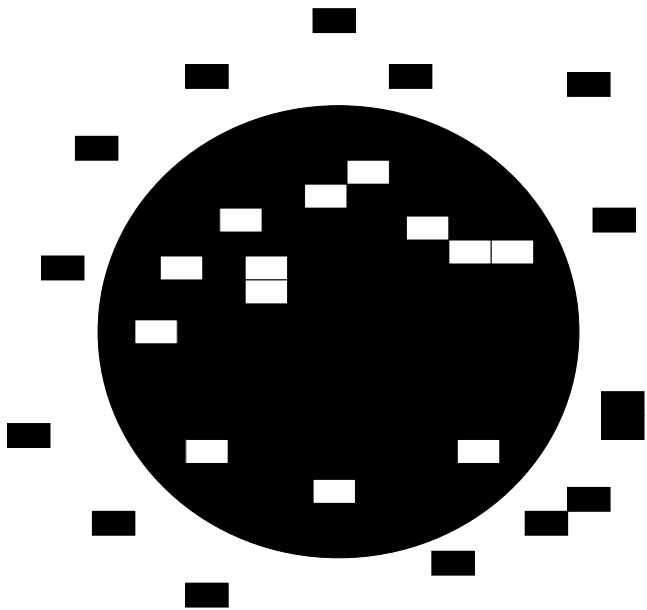}
 }
\caption{\sl Checkerboard parameterization. The white rectangles
inside the circle represent the regions $H_j$ in \eq{checkerboard},
while the black rectangles outside the circle denote the regions
$P_i$. A small number of isolated cells represents
giant gravitons in $S^5$ or in $AdS_5$. When the number of cells
is large, each additional cell (black or white) can be regarded
as a D3 brane in an arbitrary background LLM geometry defined by
the rest of the pattern.}
\label{checkerboard.fig}
\end{figure}
\be
u(x_1,x_2) = u_0(x_1,x_2) - \sum_{j=1}^m \chi_{H_j}(x_1, x_2)
+ \sum_{i=1}^n \chi_{P_i}(x_1, x_2)
\label{checkerboard}
\ee
Here $u_0$ represents a filled circle of radius $r_0$:
\be
u_0 = \theta(r_0 - r)
\label{u0}
\ee
and the regions $H_j, P_i$ are non-intersecting rectangular
cells, with $H$'s (holes) inside the circle of radius $r_0$ and 
$P$'s outside the circle.  

The constraint \eq{quad-const} is obviously satisfied. The other
constraint \eq{int-const} can also be easily satisfied, by choosing
the area of each of the cells $H_j$ or $P_i$ to be integral (in units
of $2\pi \hbar$) and by choosing the radius $r_0$ in \eq{u0} so as to
keep the total area equal to $N$. Clearly the minimum area of the
cells $H_j$ or $P_i$ is $2\pi \hbar$. In the limit of a large number
of such cells, arbitrarily scattered, we can recover a rather 
general\footnote{See footnote \ref{tiling}.} representation of the type
\eq{u-char}, subject to \eq{quad-const} and \eq{int-const}.

Thus, in \eq{checkerboard} we will choose the $H_j$ to be minimum area
cells (we will take them to be squares without loss of generality,
with each side equal to $\sqrt{2\pi \hbar} \equiv \eps $), with
centres denoted by $(x_1^j, x_2^j), j=1,...,m$. Similarly we will take
$P_i$'s to be squares of the same minimal size, with centres denoted
by $(x_1^i, x_2^i), i=1,...,n$.

The specific rectangular shape of the cells is not important for our
discussions (except for visualizing a simple tiling\footnote{
\label{tiling} The tiling is only in an approximate sense since
we will regard the cell boundaries as separated by distances $>
O(\sqrt{\hbar})$ to prevent high curvatures arising from droplets that are too
close; such inter-cell separations can be interpreted in terms of
fuzzy $u$-configurations satisfying \eq{u-star-u} in the finite
$\hbar$ theory (see Sections \ref{nc} and \ref{remarks}). Droplets
closer than this distance can be assumed to merge, leading to
``ripples''.  These are proposed in \cite{LLM} to correspond to
gravitons; we briefly explore the correspondence between gravitons and
the collective action for ripples in Appendix \ref{graviton}.}).  The
same results could be derived, e.g. by using cells with sides along
the $r$ and $\phi$ directions.

\subsection{\label{correspondence}Correspondence between checkerboard 
configurations and IIB geometries}

The correspondence with  IIB geometries, following 
\cite{LLM}, is described below:  

(a) When there are no $H$'s or $P$'s, the circle of radius $r_0$
represents \ads, where $r_0$ is given by \eq{valu-r0}.

(b) A configuration \eq{checkerboard} with a small number of
non-intersecting minimum-area cells $P_i$ and $ H_j$ represents giant
gravitons wrapping the three-spheres of $AdS_5$ or $S^5$, so that
the background configuration is essentially the same as
in the case (a) above.  The cell
$P_i$ will represent the $i$-th giant graviton extending in $AdS_5$
(such giant gravitons are called ``dual giant gravitons''
\cite{Grisaru:2000zn,Hashimoto:2000zp}).  The centre of mass of the
giant graviton will be identified with the centre $(x_1^i, x_2^i)$ of
the cell $P_i$. Similarly, the cell $H_j$ will represent the $j$-th
giant graviton extending in $S^5$ \cite{McGreevy:2000cw}. The centre
of mass of the giant graviton will be identified with the centre
$(x_1^j, x_2^j)$ of the cell $H_j$.

(c) A single minimum-area cell $H_j$ (hole) inside the filled part of a
generic $u$-configuration (representing an arbitrary LLM geometry) will
be identified as a D3-brane wrapping the three-sphere $\tilde S^3$ of
that geometry\footnote{We will not consider collective excitations
corresponding to multiple D3 branes, except to remark that 
two D3 branes which are classically on top of each other are
described in \cite{LLM} as a spread-out $u$ configuration occupying
twice the area, to be consistent with the constraints
\eq{quad-const} and \eq{int-const}. This accords  with
the fermionic description \eq{z-nc} or \eq{z-f} which we
ultimately arrive at.} (see \eq{giant-arbit}).

(d) Similarly, a minimum area cell $P_i$ in the unfilled part of an
arbitrary $u$-configuration will be identified as a D3-brane wrapping
the three-sphere $S^3$ of the corresponding geometry (see
\eq{giant-arbit-1}).

\subsection{\label{recipe}Recipe for the collective coordinate action}

We will derive the collective coordinate action\footnote{An
independent derivation, more directly from supergravity, based on
Kirillov's symplectic form, is briefly sketched in Section
\ref{remarks} (see point (2) and the references therein for details.)}
based on the above correspondences.  For example, for configurations
(b), the collective coordinate action for the $u$-fluctuation
represented by a cell $H_j$ or $P_i$ will be identified with the
action of the corresponding giant (or dual giant) graviton, subject to
the half-BPS constraint.  Similarly, for configurations (c) and (d),
the collective coordinate action will be identified with the action of
the corresponding D3-branes in an arbitrary LLM geometry, subject to
the half-BPS conditions.

To describe our method, let us consider the example of the case (c),
where we create a `hole' (a white pixel) at the position $(\bar x_1,
\bar x_2)$.  This changes the initial $u$-configuration
from an arbitrary initial value  $u_0$ to  $ u_0 - \delta u$
(where $\delta u$ is given by \eq{single-u} for a
rectangular hole). As mentioned above, this deformation $\delta u$
should be identified with a BPS D3 brane which wraps the 3-sphere
$\tilde S^3$ of the LLM geometry $u_0$ and is located
at $(\bar x_1, \bar x_2)$. The collective coordinate action $S[u]$
that we are looking for should, therefore, satisfy the property that
$\delta S = S[u_0 - \delta u] - S[u_0]$ should be identical to the action
$ S_{D3}^{BPS} $ (DBI + CS) of the above-mentioned  D3 brane
\footnote{\label{compensating} Note that 
 the configuration $u_0 - \delta
u$ does not preserve the area constraint \eq{int-const}. So we must
create another deformation $+ \delta u'$ by adding a ``particle'', or
inflating the periphery of one of the droplets comprising $u_0$. In principle
$\delta S$ could depend on the choice of  $+ \delta u'$; however,
it is easy to show that the effect is subleading in $1/N$ and
we will ignore it. This is consistent with the fact that the 
action $S[u]$ we will arrive at agrees with the  fermion action
in the semiclassical limit. We will discuss this further in
Section 7, point (6).}.

Similar considerations apply to the case (d), where one adds a
`particle' (a black pixel) at the position $(\bar x_1, \bar x_2)$ so
that $u_0 \to u_0 + \delta u$, with $\delta u$ given by \eq{single-u}.  In
this case one demands that $S[u]$ should satisfy the property that the
change in $S[u]$ should be equal to the action of a half-BPS D3-brane
at that position, wrapping $S^3$. The case (b) is of course simpler
where the background geometry is \ads\ and the D3-branes are the usual
giant or dual giant gravitons.

With the above understanding of terms, the classical action
$S[u]$ should satisfy the property 
\be
\delta S = S_{D3}^{BPS} 
\label{main-eqn}
\ee  
for an arbitrary choice of the fluctuation $\pm \delta u$,
around any background $u_0$.

We will find that such an $S[u]$ indeed exists (same as the one obtained
using the Kirillov form, Section \ref{remarks}).
  
Besides a classical action $S[u]$, we
will also find a measure $D[u]$ such that the measure for the
fluctuation $D[\delta u]_{\tilde u_0}$ agrees with the path integral
measure of the D3-brane dynamics.

Note that we are making the identification of the D3 brane
degrees of freedom with the collective coordinates of the
supergravity background. We are assuming this, as in \cite{LLM}.
This is similar in spirit with the original identification
by Polchinski \cite{Polchinski-D,Rey-collective} of 
Dirichlet branes as collective coordinates of supergravity
backgrounds carrying Ramond-Ramond charges.
 
We will discuss a more first principles approach in a later
section (Section \ref{remarks}).

Let us now consider, in turn, the D3-branes corresponding to
configurations (b), (c) and (d) of Section \ref{correspondence}.

\subsection{Single giant graviton in \ads}

In this and the next subsections we will describe the calculation
of the right hand side of \eq{main-eqn} in the cases (b), (c) and (d)
respectively. In Section 4 the action $S[u]$ and the calculation of
$\delta S$ in the left hand side of \eq{main-eqn} will be discussed.

We will first consider a giant graviton extending in $S^5$
\cite{McGreevy:2000cw}. As discussed above, this corresponds to a 
hole $H$ with each side equal to $\eps = \sqrt{2 \pi \hbar}$. We will
denote the centre of $H$ as $(\bar x_1, \bar x_2)$. The change in the
$u$-function corresponding to creation of the hole is $-\delta u$
where
\bea
\delta u &&= \chi_{\bar x_1, \bar x_2}(x_1, x_2) \nn
{}&&\equiv 
\theta(\bar x_1 + \eps/2 - x_1)
\theta(-\bar x_1 + \eps/2 + x_1) \theta(\bar x_2 + \eps/2 - x_2)
\theta(-\bar x_2 + \eps/2 + x_2)
\nn
\label{single-u}
\eea

We will now discuss the calculation of the
right hand side of \eq{main-eqn}, namely the giant graviton
action. Half-BPS configurations of a giant graviton extending in 
$S^5$ of \ads have been discussed in \cite{DJM}. The giant graviton
is a D3-brane with the embedding (in static gauge) 
\be
t = \tau, \theta = \theta(\tau), \tilde\phi = \tilde\phi(\tau),
\tilde\Omega_i= \sigma_i, \rho =0
\label{giant-embed}
\ee
where we have used global coordinates of \ads, defined by the metric
\be
ds^2 = r_0\Big[
-\cosh^2\rho~ dt^2 + d\rho^2 + \sinh^2\rho~ d\Omega_3^2
+ \cos^2\theta ~d\pht^2 + d\theta^2 + \sin^2\theta~ d \tilde \Omega_3^2 
\Big]
\label{global}
\ee
Here 
\be
r_0^2 = R_{AdS}^4 = 4\pi N l_p^4 = 4 \pi N g_s \alpha'^2
\label{valu-r0}
\ee 
The relation to the LLM coordinates   is
\bea
r &&= r_0 \cosh\rho \cos\theta
\nn
y &&= r_0 \sinh\rho \sin\theta
\label{coord-trans}
\eea
and 
\be
\phi = \tilde\phi + t
\label{phi-trans}
\ee 
For $y=0$, we have 
\be
r = r_0 \cos\theta
\label{r-for-giant}
\ee
We have used the notation 
$(r,\phi)$ as polar coordinates for the $(x_1,x_2)$ plane. 
The D3 brane action is given by \footnote{$\tilde\phi$ here is
$-\phi$ of \cite{DJM}.}
\be
S= N \int d\tau\ \Big[-\sin^3\theta\sqrt{1 - \cos^2\theta\dot\pht^2
- \dot\theta^2} - \sin^4\theta~\dot\pht\Big]
\label{giant-action}
\ee
The factor $N$ in front arises as 
\be
N = T_3 \omega_3 r_0^2,
\label{genesis-of-n}
\ee 
where
$T_3 = 1/(8\pi^3 \alpha'^2 g_s)$ is the D3-brane tension,
$\omega_3 = 2\pi^2$ is the volume of the unit $S^3$ and $r_0^2$
is given in \eq{valu-r0}.

The configuration space of the giant graviton is given by
$\theta(\tau), \pht(\tau)$.
 This corresponds to a
four-dimensional phase space $\theta(\tau), p_\theta(\tau),
\tilde\phi(\tau), p_\pht(\tau)$. 
It is easy to see that for BPS configurations
we must have \cite{DJM}
\be
\dot\theta =0, \dot{\tilde \phi}= -1
\label{vel-const}
\ee
or, equivalently,
\be
p_\theta=0, p_{\tilde\phi}= - N \sin^2\theta
\label{mom-const}
\ee
In \cite{DJM} the BPS constraints 
\eq{mom-const} were imposed as
Dirac constraints on the four dimensional phase space.
The result was a two dimensional phase space which
could be coordinatized by $\theta,\tilde\phi$ which
satisfied the following Dirac bracket:
\be
\{ \theta, \pht \}_{DB} = {1\over 2 N \sin\theta \cos\theta},
\quad {\rm or}\ \{\sin^2 \theta, \phi \}_{DB} = 1/N
\ee
The Hamiltonian in the reduced phase space is given by
\footnote{\label{moving}If we use the ``moving coordinate'' $\phi$, the 
Hamiltonian becomes $H = \tilde H + p_\pht = \tilde H + p_\phi =0$.
This is a reflection of the relation 
$ \del/\del t|_{\phi} = \del/\del t|_{\pht} + \del/\del \pht|_{t} $.
See also remarks below equation \eq{x1-x2-pht}.
}
\be
\tilde H = - p_\pht = N \sin^2\theta
\label{h-red}
\ee
Another way of stating the above result is that the
unconstrained  path integral
for the system
\be
Z_{full}= \int \D\theta(\tau)\D p_\theta(\tau)\D\pht(\tau)
\D p_\pht(\tau) \exp
\Big[i\int d\tau \Big(\dot\pht p_\pht + \dot\theta p_\theta - H_{full}\Big) 
\Big]
\label{z-full}
\ee
reduces, under the BPS constraints, to the following path
integral
\be
Z_{BPS}= \int \D[\sin^2\theta(\tau)]\D[\pht(\tau)]
\exp\Big[i \int d\tau \left( -N \sin^2\theta\dot\pht - \tilde H \right)\Big]
\label{path-for-giant}
\ee
where $\tilde H$ is given by \eq{h-red}.
We will show in Section \ref{action} how the above functional
integral can be written in terms of the $u$-variable
in the sense of Section \ref{recipe} (in
particular \eq{main-eqn}).

\gap{2}

The treatment of the dual giant graviton, extending into $AdS_5$
\cite{Grisaru:2000zn,Hashimoto:2000zp} of \ads, is very similar. The
corresponding $u$-configuration consists of a single cell $P_i$ outside
of $u_0$. Thus, 
\be
u(x_1, x_2) = u_0 + \delta u
\label{single-u-1}
\ee
where $\delta u$ is again given by the expression in
\eq{single-u}.

The D3 brane embedding for the dual giant graviton  is
\be
t=\tau, \rho=\rho(\tau),\pht= \pht(\tau),\Omega_i=\sigma_i,\theta=0
\ee
For this embedding, $r$ gets related to $\rho$ as follows:
\be
r=r_0 \cosh\rho
\label{r-for-dual}
\ee
The BPS constraints are:
\be
p_\rho=0, p_\pht = -N \sinh^2 \rho
\ee
The constrained path integral (the analog of \eq{path-for-giant})
now is 
\bea
Z_{BPS} &&= \int \D[\sinh^2\rho(\tau)]\D[\pht(\tau)]
\exp\Big[i\int d\tau \left(-N \sinh^2\rho\ \dot\pht - 
\tilde H \right)\Big]
\nn
\tilde H &&= - p_\pht =  N \sinh^2 \rho
\label{path-for-giant-1}
\eea
We will show in Section \ref{action} that this is also
a special case of the same $u$-path integral
as the earlier example was.

\subsection{\label{case-c} D3 brane in arbitrary LLM geometry}

Let us first consider configuration (c) of Section
\ref{correspondence}, where we have a single cell $H$ (hole) inside a
filled (black) region of an arbitrary $u$-configuration,
which we will write as 
(see Section \ref{recipe})
\be
u(x_1, x_2)=  u_0 -  \delta u 
\label{single-u-arbit}
\ee
where $\delta u$ is again as in \eq{single-u}, but
$u_0$ represents an arbitrary background $u$-configuration,
satisfying the constraints \eq{int-const}, \eq{quad-const}.
We will ignore here the area-compensating change $\delta u'$ as discussed
in footnote \ref{compensating}.

The D3 brane corresponding to the fluctuation
\eq{single-u-arbit} is described by the following embedding
(using the LLM coordinates, see \eq{llm-metric}):
\be
t = \tau, x_1 = \bar x_1(\tau), x_2 = 
\bar x_2(\tau), y=0, \tilde \Omega_m = \sigma_m, ~ m=1,2,3
\label{giant-arbit}
\ee
Let us discuss the geometry corresponding to $u_0$.
Recall that the LLM metric is of the form \cite{LLM}
\be
ds^2 = g_{tt} (dt + V_i dx_i)^2 + g_{yy}(dx_i dx_i + dy^2)
+ g_{\Omega\Omega}~ d\Omega_3^2 +
g_{\tilde\Omega \tilde\Omega}~d\tilde\Omega_3^2
\label{llm-metric}
\ee
where $d\Omega_3^2, d\tilde\Omega_3^2$ represent metric
on two unit 3-spheres $S^3$ and $\tilde S^3$ respectively
(the two 3-spheres
are distinguished by the fact that $S^3$ has vanishing radius
in the $u=1$ region of the $\vec x$-plane (see
\eq{black}), whereas $\tilde S^3$
 has vanishing radius
in the $u=0$ region of the $\vec x$-plane (see
\eq{white}).
The parts of the metric and RR background which are
important for us are near $y=0$:  
\bea 
u_0 &&= 1- y^2 f 
\nn 
V_i && = v_i, ~i=1,2 
\nn 
-g_{tt} &&= 1/g_{yy} = f^{-1/2} 
\nn
g_{\Omega\Omega} &&= y^2 \sqrt{f} 
\nn 
g_{\tilde\Omega \tilde\Omega} &&=f^{-1/2} 
\nn 
B_t &&= - {1\over 4} y^4 f 
\nn 
\tilde B_t && = -{1\over 4f} 
\nn 
d \hat B &&= - {1\over 4}y^3 *_3 df 
\nn 
d{\tilde {\hat B}} 
&& = -\frac12 dx_1 \wedge dx_2 = -\frac14 d (x_1~dx_2 - x_2~dx_1) 
\label{black}
\eea 
Here $*_3$ is the flat space epsilon symbol in the three
dimensions parameterized by $y,x_1,x_2$.
All expressions on the right hand sides are understood to be multiplied by
$(1 + O(y^2))$. $f(x_1,x_2), v_i(x_1,x_2)$ are both obtainable from
$u_0(x_1,x_2)$. Explicitly,
\bea
f(\vec x) &&= {\rm Limit}_{y \to 0}[\frac1{y^2} - \frac1\pi
\int_{D}\frac{d^2\vec x'}{[(\vec x - \vec x')^2 + y^2]^2}]
\nn
v_i(\vec x) && =
{\eps_{ij}\over 2\pi}\oint_{\del D} \frac{dx'_j}{ (\vec
x - \vec x')^{2}}
\label{def-f-v}
\eea
Here $D$ denotes the support of $u$.
The limit for $f$ is well-defined since the explicit $1/y^2$ cancels
with a $1/y^2$ coming from the $\vec x = \vec x'$ region
of the integral. It is easy to calculate explicit forms for $f$,
for example, for ring configurations of $\tilde u_0$. 

Under the approximations \eq{black} the  metric
and the RR 4-form field are given, upto $(1 + O(y^2))$, by
\bea
ds^2 && = [-( dt + v_i dx_i)^2 + f(dx_1^2 + dx_2^2
+ d\vec y^2) +  d\tilde\Omega^2]/\sqrt{f}
\nn
C^{(4)} && = -\frac14 \Big[ \frac{dt + v_i dx_i}f + r^2 d\phi \Big]
\wedge d^3\tilde \Omega
\label{fields}
\eea
where $d\vec y^2= dy^2 + y^2 d\Omega^2$, and $d^3\tilde \Omega$
is the volume form of the three-sphere $\tilde S^3$ (see
\eq{llm-metric}).

The D3 brane action is given by\footnote{Note the appearance in the
second line of the
$\hbar$ of \eq{int-const},\eq{hbar-llm} through the
equality $T_3 w_3 = N/r_0^2 \equiv 1/(2 \hbar)$, cf.
\eq{genesis-of-n}.} (dropping the bar's on $x_i(t)$
in \eq{giant-arbit})
\bea
S && = T_3 \omega_3 \int d\tau \Big[
-\frac1f \sqrt{(1 + v_r \dot r + v_\phi \dot \phi)^2  - 
f(\dot r^2 + r^2 \dot \phi^2)}
+ r^2 \dot \phi + \frac1f (1 + v_r \dot r + v_\phi \dot \phi) \Big]  
\nn
&&= \frac{1}{2 \hbar} \int dt \Big[
-\frac1f \sqrt{(1 + v_r \dot r + v_\phi (\dot \pht +1))^2  - 
f(\dot r^2 + r^2 (\dot \pht +1)^2)} \Big.
\nn
&&~~~~~~~~~~~~~~~~~
\Big.  + r^2 (\dot\pht +1) + \frac1f (1 + v_r \dot r + v_\phi(\dot \pht +1)) 
\Big]  
\label{action-arbit}
\eea
The BPS conditions can be obtained by the constraint
$\tilde H = - p_\pht$, which gives
\be
\dot\pht = -1, \dot r=0
\ee
In the $\phi, t$ coordinates
\be
\dot\phi = 0, \dot r=0
\label{vel-const-arbit}
\ee
The Hamiltonian $H$ in the LLM frame is $H=0$ (see footnote
\ref{moving}). It should be possible to derive 
these equations from an analysis
of the Killing spinor and world-volume kappa-symmetry, but
another way of seeing the validity of equations \eq{vel-const-arbit}
is that it is equivalent to time-independence of $\delta u$ in 
\eq{single-u-arbit}. Any such time-independent
$u$-configuration is half-BPS, as shown in
\cite{LLM}; indeed the half-BPS condition 
does not allow
any time-dependence of $u$. Hence \eq{vel-const-arbit} is
equivalent to the Killing spinor condition. 

The remaining analysis is similar to the case of the giant
gravitons in \ads. On the constrained surface \eq{vel-const-arbit}
we have
\be
p_r =0, p_\pht = \frac1{2 \hbar} r^2
\ee
The Hamiltonian is given by 
\be
\tilde H= - p_\pht = -  \frac1{2 \hbar} r^2,
\label{ham-giant-arbit}
\ee
the negative sign reflecting the energy of a hole. 

The constrained path integral, the analog of
\eq{path-for-giant}, now becomes
\bea
Z_{BPS} &&= \int \D[r^2(\tau)]\D[\pht(\tau)]
\exp[i S_{BPS}]
\nn
S_{BPS} &&= \int d\tau \left( \frac1{2 \hbar} r^2 \dot\pht - \tilde H 
\right)
\label{path-for-giant-arbit}
\eea
where $\tilde H$ is given by \eq{ham-giant-arbit}. To compare 
with \eq{path-for-giant}, note that on \eq{r-for-giant}
$r^2/(2\hbar)= N \cos^2\theta = N - N \sin^2\theta$. The extra $N$
is explained in the paragraph following \eq{delta-s-ham}.

\gap{2}

Let us now consider configuration (d), where we have a single 
(black) cell $P$ 
in a white   region  of an arbitrary $u$-configuration. The full
u-configuration, including contribution from $P$ is given by 
\be
u(x_1, x_2) =  u_0 + \delta u
\label{single-u-arbit-2}
\ee
where $\delta u$ is given by \eq{single-u}. 

As in \eq{black}, the important parts of the metric and RR background are near
$y=0$. These are now given by
\bea
u_0 &&= y^2 f
\nn
V_i && = v_i 
\nn
-g_{tt} &&= 1/g_{yy} = f^{-1/2}
\nn
g_{\Omega\Omega} &&= f^{-1/2}
\nn
g_{\tilde\Omega  \tilde\Omega} &&= y^2 \sqrt{f}
\nn
B_t &&=  - {1\over 4f}
\nn
\tilde B_t && =- {1\over 4} y^4 f
\nn
d \hat B &&= 
\frac12 dx_1 \wedge dx_2 = \frac14 d (x_1~ dx_2 - x_2~ dx_1)
\nn
d B && =  {1\over 4}y^3 * df
\label{white}
\eea
All expressions on the right hand sides are understood to be 
multiplied by
$(1 + O(y^2))$. $v_i$ are again given by \eq{def-f-v},
while $f= (1/\pi) \int_D d^2 \vec x'~(\vec x - \vec x')^{-4}$.

The metric and the RR form are given by  
\bea
ds^2 && = [-( dt + v_i dx_i)^2 + f (dx_1^2 + dx_2^2
+ d\vec y^2) + d \Omega^2]/\sqrt{f}
\nn
C^{(4)} && = -\frac14 \Big[ \frac{dt + v_i dx_i}f - r^2 d\phi \Big]
\wedge d^3\Omega
\label{fields-1}
\eea
where $d\vec y^2= dy^2 + y^2 d\tilde \Omega^2$ and $d^3\Omega$
represents the volume-form on $S^3$ (see \eq{llm-metric}). 

Let us consider the D3 brane represented by $\delta u$ in
\eq{single-u-arbit-2}.  Its embedding is given by
(using, again, the LLM coordinates of \eq{llm-metric})
\be
t=\tau, x_1= \bar x_1(\tau), x_2= \bar x_2(\tau), y=0,
\Omega_m = \sigma_m, ~m=1,2,3
\label{giant-arbit-1}
\ee
The D3 brane action, analogous to \eq{action-arbit}, is
given by  (dropping the bar's on $x_i(\tau)$)
\bea
S && = T_3 \omega_3 \int d\tau \Big[
-\frac1f \sqrt{(1 + v_r \dot r + v_\phi \dot \phi)^2  - 
f(\dot r^2 + r^2 \dot \phi^2)}
- r^2 \dot \phi + \frac1f (1 + v_r \dot r + v_\phi \dot \phi) \Big]  
\nn
&&= \frac{1}{2 \hbar} \int d\tau \Big[
-\frac1f \sqrt{(1 + v_r \dot r + v_\phi (\dot \pht +1))^2  - 
f(\dot r^2 + r^2 (\dot \pht +1)^2)} \Big.
\nn
&&~~~~~~~~~~~~~~~~~
\Big.  - r^2 (\dot\pht +1) + \frac1f (1 + v_r \dot r + v_\phi(\dot \pht +1)) 
\Big]  
\label{action-arbit-2}
\eea
The BPS condition $H= - p_\pht$, once again equivalent to
$\dot \pht = -1, \dot r =0$, implies that the BPS dynamics
is described by the path integral (analog of \eq{path-for-giant-arbit})
\bea
Z_{BPS} &&= \int \D[r^2(\tau)]\D[\pht(\tau)]
\exp[i S_{BPS}]
\nn
S_{BPS} &&=  \int d\tau \left(
-\frac{r^2}{2\hbar} \dot\pht - \tilde H \right)
\nn
\tilde H  && = - p_\phi = \frac1{2\hbar} r^2
\label{path-for-giant-arbit-1}
\eea
Note that the Hamiltonian for the filled cell is
positive this time. For comparison with \eq{path-for-giant-1}, remarks
similar to the ones below \eq{path-for-giant-arbit} apply here as
well (note that according to \eq{r-for-dual} $r^2/(2\hbar) = N \cosh^2\rho
= N + N \sinh^2 \rho$).

\section{\label{action}Collective coordinate action}

We will now show that all the path integrals 
\eq{path-for-giant},\eq{path-for-giant-1},\eq{path-for-giant-arbit}
and \eq{path-for-giant-arbit-1} are 
equivalent to the  following path integral in terms
of the $u$-variable:
\bea
Z &&= \int \D\u \exp[i S_{BPS}]
\nn
S_{BPS}
&& = \int {dx_1 dx_2 \over 2\pi \hbar}
\hbar \int_{\tilde \Sigma} d\tau\ 
ds\ \u(x_1,x_2,\tau,s)\{\del_\tau \u, \del_s \u \}_{PB} -  \int_\Sigma d\tau
\tilde H 
\label{u-action}
\nn
\tilde H &&= \int {dx_1 dx_2 \over 2\pi \hbar}
u(x_1,x_2,\tau) \frac{x_1^2 + x_2^2}{2 \hbar}
\label{path-for-u}
\eea
Here $\Sigma$ denotes a 
curve $\tau\! \mapsto\! u(x_1,x_2,\tau)$
in the $u$-configuration space and  $\tilde \Sigma$ 
denotes a one-parameter  extension of $\Sigma$
to the map $(\tau,s)\! \mapsto \! u(x_1,x_2,\tau,s), s < s_0,$ such that
$u(x_1,x_2,\tau,s_0)= u(x_1,x_2,\tau)$. Although in order
to write the action we need to introduce the $s$-extension,
it can be easily shown that the
extension does not affect the path integral as long as the boundary
value (at $s=s_0$) remains $u(x_1,x_2,\tau)$ (this
follows from the fact that the symplectic form 
appearing in \eq{u-action} is closed). In this and the
following section we use 
\be
(x_1, x_2) = (r \cos\pht, r \sin\pht)
\label{x1-x2-pht}
\ee
(see eq. \eq{phi-trans}). The $\pht$ coordinate, rather than $\phi$,
is the more natural angle to use for comparison with the
boundary theory, because, e.g., the time-derivative in the boundary
theory is the operator $\del/\del t|_\pht$ appearing in 
footnote \ref{moving}. In terms of $(r,\phi)$ coordinates
the Hamiltonian is zero (see footnote \ref{moving}).  

The notation $\{ \}_{PB}$ is defined here as
\[
\{ f, g \}_{PB} \equiv \frac{\del f}{\del x_1}\frac{\del g}{\del x_2}
- \frac{\del g}{\del x_1}\frac{\del f}{\del x_2}
\]

We will see later (see Section \ref{remarks} and references
therein) that the action \eq{u-action}
is Kirillov's coadjoint orbit action for the group of area-preserving
diffeomorphisms. 

The measure $Du$, described in Sections \ref{measure}
and \ref{remarks}, incorporates the constraints \eq{quad-const}
and \eq{int-const}.
The equation of motion
for $u(x_1,x_2,\tau)$ that follows from \eq{path-for-u} is
(see \cite{DMW-classical,DMW-nonrel}):
\be
\del_\tau u - (x_1 \del_2 - x_2 \del_1) u=0
\label{eom-u}
\ee

\subsection{Action}

We will show that the action \eq{u-action} gives rise to
the various D3-brane actions in
\eq{path-for-giant},\eq{path-for-giant-1},\eq{path-for-giant-arbit}
and \eq{path-for-giant-arbit-1}  in the sense of 
\eq{main-eqn}. 
Consider, for example,  configuration (d), \eq{single-u-arbit-2},
\eq{giant-arbit-1}.
It is easy to see that if  $\delta u$ does not
intersect with $u_0$, 
then the left hand side of \eq{main-eqn}
is given by local properties of the cell $\delta u$, viz.
$\delta S[u] = S[\delta u]$. Thus we get 
\bea
\delta S &&= \delta S_{kin} - \delta S_{ham}
\nn
\delta S_{kin} && = \int {dx_1 dx_2 \over 2\pi \hbar}
\hbar \int_{\tilde \Sigma} d\tau 
ds\ \delta u\{\del_\tau \delta u, \del_s \delta u \}_{PB} 
\nn
\delta S_{ham} &&= \int d\tau \int {dx_1 dx_2 \over 2\pi \hbar}
\delta u(x_1,x_2,\tau) \frac{x_1^2 + x_2^2}{2 \hbar}
\label{action-for-delta-u}
\eea
We need to show that the above action is equal to the
action $S_{BPS}$ appearing in  \eq{path-for-giant-arbit-1}.

Let us consider first the Hamiltonian term:
\bea
\delta S_{ham} && = 
\int d\tau \langle \frac{x_1^2 + x_2^2}{2 \hbar} \rangle 
 \int {dx_1 dx_2 \over 2\pi \hbar}
\delta u(x_1,x_2,\tau) 
\nn
&&= \int d\tau \frac{{\bar x_1}^2 + {\bar x_2}^2}{2 \hbar}
\nn
&& = \int d \tau \frac{r^2}{2 \hbar}
\label{delta-s-ham}
\eea
which matches with the Hamiltonian term in \eq{path-for-giant-arbit-1}. 
In the first step we have taken the integrand out of the cell $\delta u$
since its size is small, in the second step we have used the fact that
$\delta u$ has area $2\pi \hbar$ and also equated the average values of
$x_1, x_2$ with the coordinates of the centre of mass 
$\bar x_1, \bar x_2$ (see \eq{giant-arbit-1}) 
which satisfies ${\bar x_1}^2 + {\bar x_2}^2 = r^2$.

The analysis of the Hamiltonian term for configuration (c)
[\eq{single-u-arbit},\eq{giant-arbit},\eq{path-for-giant-arbit}] is
similar. It is interesting to note that in the special cases
\eq{path-for-giant} and \eq{path-for-giant-1} the Hamiltonian by
convention measures the energy of the fluctuation $\delta u$ together
with that of a compensating fluctuation $\delta u'$
(see footnote \ref{compensating}) defined by
adjusting the radius $r_0$ (this, again, corresponds to a choice of
gauge for $C^{(4)}$ different from that in \eq{fields},
\eq{fields-1}). Thus, e.g. the energy \eq{h-red} includes the energy
of the hole $-N \cos^2 \theta$ as well as the energy $+N$ of the
compensating outer circular strip $+\delta u'$, extending
between $r_0$ and $r_0 + \delta r_0$ such that the latter
radius has an area $N+1$. In the generic case it is more natural
to keep the two effects separate, which is possible to do in 
the semiclassical limit.

The analysis of the kinetic term $\delta S_{kin}$
is more complicated and is presented in Appendix A. It is,
however, somewhat simpler to match the equation of motion that follows
from \eq{action-for-delta-u} with the equations of motion following
from \eq{path-for-giant-arbit-1}. The latter are
\be
\dot{\bar x_1}= \bar x_2, \dot{\bar x_2} = - \bar x_1
\label{sho}
\ee
The equation of motion following from the action
\eq{action-for-delta-u} can be read off from
\eq{eom-u} and is given by
\be
\dot{\delta u} - (x_1 \del_2 - x_2 \del_1)\delta u=0
\label{eom-delta-u}
\ee
Using the expression \eq{single-u} for $\delta u$, 
one can show that \eq{eom-delta-u} is satisfied to leading
order in $\hbar$, provided \eq{sho} is valid. 

\subsection{\label{measure}Measure}

The measure $\D u$ is defined as the group-invariant measure where $u$
is parameterized as an orbit of some specific field configuration
$u_0$ under the group of area-preserving diffeomorphisms (see
\cite{DMW-classical,DMW-nonrel} and Section \ref{remarks}). The
reference configuration $u_0$ satisfies $u_0^2 = u_0$ and 
$\int dx_1 dx_2~ u_0/(2\pi\hbar)= N$ so that 
the measure $Du$ incorporates the two
constraints \eq{quad-const} and \eq{int-const}.

When $g$ acts on $\delta u$ (see \eq{single-u}), the action gets
transmitted to the centres of mass of $\delta u$ as a canonical
transformation on $\bar x_1, \bar x_2$ (cf. \eq{group-action}). The
invariant measure under canonical transformations is the one already
used in \eq{path-for-giant-arbit-1}.  We find, therefore, that the
measures also agree.

\section{\label{nc} Equivalence to Fermion path integral}

Ref. \cite{DMW-path} discussed 
the  following path integral which
represented a path integral for the phase space density
$u(q,p,t)$ for free fermions moving in one
dimension under a Hamiltonian $h(q,p)$
\bea
Z_{NC} &&= \int [Du(q,p,t)]_{u_0} \exp[i S[u]]
\nn
S[u]
&& = \int {dq~dp \over 2\pi \hbar}
\hbar \int_{\tilde \Sigma} dt~ ds~ 
\u(q,p,t,s) * \{\del_t \u, \del_s \u \}_{MB} -  
\int_\Sigma dt
\tilde H 
\nn
\tilde H &&= \int {dq~dp \over 2\pi \hbar}
u(q,p,t)* {h(q,p)\over \hbar}
\label{z-nc} 
\eea  
For free fermions moving in a harmonic oscillator potential
\be
h(q,p) = {p^2 + q^2\over 2}
\label{harmonic}
\ee
The star product in \eq{z-nc} is defined as
\be
a * b (q,p) 
= \left[\exp\Big({i\hbar\over 2}\left(
\del_{q}\del_{p'} -  \del_{q'}\del_{p}\right)\Big)
\left(a (q,p) b(q',p') \right)\right]_{q'=q,p'=p}
\label{star}
\ee
The Moyal Bracket is defined as
\be
\{ a, b\}_{MB} = \frac{a * b - b * a}{i\hbar}
\label{Moyal}
\ee
The measure $Du$ is defined as the group-invariant measure
under the symmetry group $W_\infty$ of the 
fermion configuration space \cite{DMW-path,DMW-nonrel}.
The space of $u$'s is the $W_\infty$ orbit of a
reference configuration $u_0$ which we can take
to be the expectation value of
the Wigner phase space distribution \eq{wigner}
in the Filled Fermi sea.
The measure incorporates the constraint
\be
u * u = u
\label{u-star-u}
\ee
and
\be
\int \frac{dq dp}{2\pi \hbar} u = N
\label{int-const2}
\ee
The operator definition of the Wigner distribution $\hat u(q,p,t)$
is given in \eq{wigner}.

The equation of
motion following from this path integral is
\bea
\del_t u(q,p, t) && = \{h(q,p), u(q,p, t)\}_{MB}
\nn
{}&& = \{h(q,p), u(q,p, t)\}_{PB}
\nn
{}&& = (q\del_p - p \del_q) u(q,p,t)
\label{eom-old}
\eea
$h(q,p)$ is the single particle Hamiltonian appearing
in \eq{z-nc}.
The second step follows for any quadratic Hamiltonian. 
For the  $c=1$ matrix model, one takes $h = 
(p^2 - q^2)/2$, but the analysis in \cite{DMW-path} is true for
any Hamiltonian and in particular for $h =(p^2 + q^2)/2 $. 
The third line follows from this latter Hamiltonian.
Although the equation of motion \eq{eom-old} coincides with its classical
limit \eq{eom-u}, the finite $\hbar$ dynamics differs significantly 
from its classical limit because the constraint \eq{u-star-u} involves
star products, involving fuzzy solutions for $u$ 
\cite{DMW-nonrel,DMW-instanton}, unlike the constraint
\eq{quad-const} whose solutions are characteristic functions
\eq{u-char}. This is discussed further in the next two sections.

In \cite{DMW-path} it was shown that \eq{z-nc} is exactly
equal to a  path integral for $N$ free fermions moving in a simple
harmonic oscillator potential, defined as follows:
\bea
Z_{NC}= Z_F &&= \int D[\Psi]_{\ket{F_0}} \exp[i S_F/\hbar]
\nn
S_F &&= \int dt\ dx \ [\Psi^\dagger(x,t)
(i \hbar \del_t - h(x, \del_x)) \Psi(x,t)]
\nn
h &&=\frac12(-\hbar^2 \frac{\del^2}{\del x^2} + x^2)
\label{z-f}
\eea  
Here $\Psi(x,t), \Psi^\dagger(x,t)$ are the second
quantized annihilation and creation operators (respectively)
for the fermions.
The subscript ${\ket{F_0}}$ in the measure implies that
the functional integral is over states 
obtained from the reference Fock space state $\ket{F_0}$
under $W_\infty$ transformations.  
These in fact span all states with the same
fermion number as $\ket{F_0}$, which we take to be $N$.

\myitem{Wigner phase space distribution}

The Wigner phase space  distribution $u(q,p,t)$ which appears in
\eq{z-nc} as a path integral variable, can be defined
as an operator (second quantized, see, e.g. 
\cite{DMW-nonrel,DMW-instanton}) 
as follows:
\be
\hat u(q,p,t) = 
\int d\eta \Psi^\dagger(q+ \eta/2,t)\Psi(q-\eta/2,t) \exp[i{p
\over \hbar}\eta ] 
\label{wigner}
\ee
Salient properties of this quantity as well those of
its expectation values in various states have been
listed in \cite{DMW-nonrel,DMW-instanton, DMW-classical}.

\myitem{The correspondence}

It is clear that \eq{path-for-u} is simply the $\hbar\to 0$ limit of
\eq{z-nc}, provided one identifies $u(x_1, x_2)$ of Section 4
with $u(q,p)$ of this section. This is the advertised
transformation of configuration  space into phase space. 
The constraints \eq{quad-const}
and \eq{int-const} also follow from \eq{u-star-u} and 
\eq{int-const2}. Note that the equation $u * u =u$
reduces to $u^2 = u$ in the semiclassical limit, a fact
which has been extensively exploited in \cite{DMW-nonrel,
DMW-instanton, DMW-classical}.

Hence the collective coordinate quantization of LLM
geometries gives rise to the $\hbar \to 0$ limit of free fermions in a
harmonic oscillator potential. This is of course what we expect from
the AdS/CFT correspondence \cite{LLM}, but we arrived at this result here
starting from D-branes in supergravity. How to elevate this result to finite
$\hbar$ remains an interesting issue. Some possible
subtleties are mentioned in the next section. In the next section
we also briefly discuss a more direct derivation of the semiclassical
correspondence from supergravity using Kirillov's symplectic
form.  

\section{\label{remarks}Remarks on collective coordinate 
method with BPS constraint}

In this section we will briefly discuss a first principles approach to
the collective coordinate quantization of half-BPS geometries without
using the D3 brane actions.

We begin by noting that the group G of  time-independent area-preserving
diffeomorphisms (SDiff) is a symmetry of the constraints \eq{int-const} and
\eq{quad-const}, as well as of the equations of motion of the type IIB
theory (since the geometries corresponding to various $u$'s
all satisfy IIB equations of motion).
The Lie Algebra $\bar G$ is the algebra of symplectic vector fields. Thus,
elements $g= 1 + X_f$ near identity of $G$,  act on a function $u(x_1,x_2)$ 
as 
\bea 
u\to u^g &&= u + X_f.u = u + \{f, u\}_{PB} 
\nn 
X_f &&=
\epsilon_{ij} {\del f\over \del x_i}{\del \over \del x_j} 
\eea 
This action can also be regarded as induced by the motion
of points on the plane under a Hamiltonian $f$:
\bea
u^g(x)&& = u(x^{g^{-1}}), {\rm where}
\nn
(x_1, x_2)^g &&\equiv (x_1 + \del f/\del x_2, x_2 - \del f/\del x_1)
\label{group-action}
\eea
Finite group elements $g \in {\rm SDiff}$ can be dealt with  by
exponentiation. 
Now, since the function $u$ completely determines the supergravity fields
(collectively denoted below as $\Phi$):
$ \Phi = \Phi[u]$, the  group $G$ of area-preserving diffeomorphisms
has a natural action on  supergravity fields:
\be
\Phi^g = \Phi[u^g]
\ee
The choice of any given function $u_0$, and the corresponding
$\Phi_0$  breaks the
symmetry $G \to H$, where $H$ denotes the subgroup generated by
functions which have zero Poisson bracket with $u_0$. 

The collective  coordinate method
\cite{Gervais:1975yg,Gervais:1976ca} consists of making a change of
variable $\Phi(t) \to \{ g(t), \tilde \Phi(t)
\equiv \Phi^{g(t)}(t) \}$, where $\tilde \Phi(t)$ represents
motion in the body-fixed frame which is over and above the
collective motion. The dynamics of the collective
coordinate  is obtained by implementing the change of variable
in the field theory functional integral.

A first principles derivation of the collective coordinate action
(without using the identification with D3 branes) would involve
implementing the above procedure in the case of IIB supergravity.
We will not attempt to do this here in detail, but give
a brief discussion:

\begin{enumerate}

\item 
Since the IIB Lagrangian is second order in time derivatives, the low
energy action for $g(t)$, is expected to be quadratic in $\dot g$
(before implementing the BPS condition).  This corresponds, for
example, to \eq{giant-action}, \eq{action-arbit} or
\eq{action-arbit-2}, which are second order in time at low
velocities. The phase space of the collective degree of freedom $g(t)$
at this stage involves $g(t)$ as well as
$\pi_g(t)$ where $ \pi_g(t)$ is the ``momentum''
for $g(t)$.

\item 
In case of the D3 brane dynamics one can explicitly see how
\eq{z-full} changes to \eq{path-for-giant} with the imposition of the
BPS constraint \eq{mom-const}. One would similarly expect that, 
if one implements the change of variable $\Phi \to
\{ g(t), \tilde \Phi \}$ in the IIB functional
integral ``in the presence of a BPS constraint'',  
the dynamics of the collective variable $g(t)$ will be
described by a first-order action and $g(t)$'s themselves would become
a phase space. The most natural such an action on a
$G$-orbit of a configuration $u_0$ is 
given by Kirillov's method of coadjoint orbits 
\cite{DMW-nonrel,DMW-classical,Kirillov}
(see, e.g  eq. (68) of \cite{DMW-classical})  
\be
S_{BPS} 
=  \int dt \langle X_t, u_0 \rangle
- \int dt \langle g^{-1} X_h g, u_0 \rangle  
\label{action-g}
\ee 
where $\langle X_f, u_0 \rangle \equiv \int \frac{dx_1 dx_2}{2\pi
\hbar} (f(x_1, x_2) u_0(x_1, x_2))$. The notation $X_t$ denotes the
Lie algebra element $g^{-1} \dot g$ and $X_h \equiv
g^{-1} h g$ denotes the $g$-transported Lie algebra element
corresponding to the   
Hamiltonian $h = (x_1^2 + x_2^2)/2$.
This action exactly coincides with \eq{u-action}
\cite{DMW-nonrel,DMW-classical}.  Indeed the measure also coincides
with the measure of \eq{u-action}. As a matter of fact 
we initially arrived at the
action \eq{u-action} by considering the Kirillov action
\cite{Mandal-talk};
from this viewpoint the D3-brane method can be viewed 
as additional evidence in support of the Kirillov action. 

\item 
Evaluation of a functional integral ``in the presence of a BPS
constraint'' involves insertion of an appropriate projection operator.
It is possible that the resulting functional integral is an index, as
in \cite{Maldacena:1999bp,Cecotti:1992qh}, which are natural tools for
counting geometries satisfying a specific number of supersymmetries.

\item
If all $\Phi[u]$'s can be generated by the collective motion
$\Phi[u^g]$, clearly the other degrees of freedom $\tilde \Phi$ are to
be omitted from the functional integral under the half-BPS
constraint. In this sense, the collective coordinate functional
integral would appear to be the entire supergravity functional
integral when subjected to the half-BPS constraint (see the next
point, however).

\item
There is an important subtlety regarding the number of connected
components of the $u$-configurations. Although SDiff acts on the LLM
geometries, it is not clear how it can change the number of connected
components of a given $u$-configuration. Of course, under the
$W_\infty$ group mentioned in Section \ref{nc}, such a transformation
can happen (a fuzzy droplet can split into two fuzzy
droplets). However, $W_\infty$ is the symmetry group of the equation
$u * u =u$ and is not naturally associated with the LLM constraint
$u^2 = u$. It remains unclear to us at the moment how to describe the
entire space of LLM geometries as the orbit of a given configuration
under a certain group $G$.

\end{enumerate}

\section{Conclusion}

In this paper we considered collective coordinate quantization of LLM
geometries identifying the function $z(x_1,x_2,0)\equiv 1/2 -
u(x_1,x_2)$ of \cite{LLM} as the collective coordinate. The explicit
form of the collective coordinate action (and measure) is derived by
identifying the collective degree of freedom as that of a D3 brane
coupled to an arbitrary LLM geometry.  The D3 brane functional
integral, subject to the BPS constraint, can be written directly in
terms of the $u$-variable. We show that the resulting functional
integral is the $\hbar\to 0$ limit of a functional integral describing
free fermions in a harmonic oscillator potential.  We discuss a first
principles approach towards derivation of the $u$-integral using the
general method of collective coordinates subject to a BPS constraint.

We note a few important points:

\begin{enumerate}

\item 
We find that supergravity configuration space becomes a phase space
(hence noncommutative, with a noncommutativity parameter given by a
certain $\hbar$), when constrained to configurations preserving a
certain number of supersymmetries. Although we found this phenomenon
in a specific case here (half-BPS IIB supergravity solutions with
$O(4) \times O(4)$ symmetry), it is clear that this phenomenon should
be generic. In particular the appearance of a first order action,
discussed in Section \ref{remarks}, is related to the fact that the
BPS equations are first order. The formalism of phase space path
integrals employed in this paper makes it rather apparent how a
configuration space path integral with second order action becomes a
phase space path integral with first order action under the imposition
of the BPS constraint.  It appears to be possible, using this, to
count supersymmetric configurations within low energy field theories
including supergravity. This observation clearly has implications for
counting entropy of supersymmetric black holes and other related
configurations. 

\item
As we mentioned in the previous section, functional integrals
preserving a certain number of supersymmetries have earlier been
treated in, for example, \cite{Maldacena:1999bp,Cecotti:1992qh}, where
the partition function is a `twisted' one involving insertion of
operators related to $(-1)^F$.  It would be interesting to see if this
is the case for half-BPS supergravity solutions treated in this
paper. One would imagine defining such path integrals in terms of
projection operators in the Hilbert space enforcing the supersymmetry
conditions; it is of interest to explore the connection between this
definition and the `twisted' partition function mentioned above.
Another related way of understanding ``BPS functional integrals''
would be to use topological twisting so that the relevant
supersymmetry operators become BRST operators and the desired path
integral becomes the normal path integral in the topological
theory.
  
\item
It is entirely possible, as in the context of the $c=1$ matrix models,
that the semiclassical collective excitation approach misses important
subtle points of the fermion theory. In the case of $c=1$ this was
discussed in great detail in \cite{Sengupta-Wadia,
Mandal:1991ua,DMW-nonrel,DMW-instanton,DMW-beta,DMW-discrete}. One
important effect missed by classical collective excitations
(corresponding to the massless `tachyons') is the unstable D0 brane of
the two-dimensional string theory
\cite{McGreevy:2003kb,Klebanov:2003km} (this viewpoint is explained in
\cite{Mandal:2003tj}).  In the present case, the semiclassical
collective excitations consist of  ripples (corresponding to
gravitons, see Appendix B) as well as D3 branes (roughly analogous to the
tachyons and D0 branes, respectively, of  two dimensional string
theory). However, we might discover other important effects related to
the non-perturbative description \eq{z-nc} possibly missed by the
semiclassical treatment of the collective excitations.

\item 
We have used D3 branes coupled to LLM geometries to find
noncommutative dynamics in the configuration space.  It is interesting to
note that in the limit of LLM geometries  which describes D3
branes in the Coulomb branch \cite{LLM}, the value of $\hbar$ scales to zero
causing the noncommutativity to disappear, as one would expect.

\item
As seen in Section 5, the phase space density action obtained from the
fermion theory has an additional degree of noncommutativity reflected
in the appearance of star products, over and above the
noncommutativity mentioned in points (1) and (4). The latter is
already evident in the semiclassical limit itself where the Moyal
brackets get reduced to Poisson brackets and reflects a phase space
structure of the classical configuration space. Clearly the former is
related to the issue of finite $\hbar$ correspondence between the
half-BPS geometries and the fermion theory. Of particular importance
is whether the generalization to the constraint $u * u = u$ (instead
of $u^2 = u$) allows some insight into $g_{st}$ effects in string
theory. Some aspects of the effect of finite $g_s$ have been discussed
at the end of the previous section. 

\item
A specific subleading $1/N$ correction briefly mentioned in this paper
is the effect of the compensating fluctuations $\delta u'$ (see
footnote \ref{compensating}). This effect is indeed calculable in the right
hand side of \eq{main-eqn} for various choices of $\delta u'$ and it
is an interesting question whether the corresponding modification
in the left hand side arises correctly by taking into account
interaction between $\delta u$ and $\delta u'$ coming from
the star product structure of $S[u]$.

\item 
Most of this paper dealt with collective excitations
identified as D3 branes. We discuss gravitons briefly
in Appendix B; it would be interesting to quantitatively
reproduce the graviton fluctuations from our collective action.
 
\end{enumerate}

\section{Acknowledgments}

I would like to thank David Berenstein, Avinash
Dhar, Jaume Gomis, Oleg Lunin, Alex Maloney, Liam McAllister, John
McGreevy, Rob Myers, Hiroshi Ooguri, Joe Polchinski, Yasuhiro Sekino,
Lenny Susskind and Spenta Wadia for discussions, Nemani Suryanarayana
for discussions and initial collaboration and 
Shahin Sheikh-Jabbari for comments.
I would like to acknowledge the  hospitality at SLAC, Stanford and
Caltech during the finishing stages of this work.

\appendix

\section{Phase space density action for a single cell}

In this Appendix we will evaluate $\delta S_{kin}$
appearing in \eq{action-for-delta-u}, with
$\delta u$ as in \eq{single-u}.
For simplicity of notation, we will denote 
\be
x_1 =q, x_2 = p
\label{simple}
\ee
Let us define 
\be
q^\pm  = \pm \bar q + \eps/2 \mp  q,  
p^\pm  = \pm \bar p + \eps/2 \mp  p
\ee
Then 
\be
\delta u(q,p) = \theta(q^+)\theta(q^-)\theta(p^+)\theta(p^-)
\ee
It is easy to calculate
\bea
\dot{\delta u}
&&= \dot{\bar p}\Big[
\delta (p^+) \theta(p^-) - \theta(p^+)\delta(p^-)\Big]
\theta(q^+) \theta(q^-)
\nn
&&+  \dot{\bar q}
\Big[\delta (q^+) \theta(q^-) - \theta(q^+)\delta(q^-)
\Big]
\theta(p^+) \theta(p^-)
\eea
and
\bea
{\delta u}'
&&= {\bar p}'
\Big[ \delta (p^+) \theta(p^-) - \theta(p^+)\delta(p^-)
\Big] 
\theta(q^+) \theta(q^-)
\nn
&&+ {\bar q}'
\Big[\delta (q^+) \theta(q^-) - \theta(q^+)\delta(q^-)
\Big]\theta(p^+) \theta(p^-)
\eea
We define the Poisson bracket 
\be
\{f,g\}_{PB} = \del_q f \del_p g
- \del_p f \del_q g
\ee
We get, after some simplification,
\bea
\delta S_{kin}
&&=  \int d\tau ds\int \frac{dq dp}{2\pi \hbar} \hbar
\delta u \{\dot{\delta u}, {\delta u}'\}_{PB} 
\nn
&&= \int \frac{3 dp\ dq}{8 \pi} \{
\delta^2(q^+) +\delta^2(q^-)\}\{\delta^2(p^+) + \delta^2(p^-)\}
\Big[ \int d\tau ds\ (\dot{\bar q}{\bar p}' -  \dot{\bar p}{\bar q}')
\Big]
\nn
&&= A
\int d\tau (- \frac{r^2}{2\hbar} \dot \pht)
\nn
A && = \frac{3\hbar}{\pi}\delta_q(0)\delta_p(0)
\label{tmp}
\eea
In the last line we have used eqs. \eq{x1-x2-pht} and
\eq{simple} and the equality 
\be
\dot{\bar q}{\bar p}' -  \dot{\bar p}{\bar q}' 
= \del_s (\bar p \dot{\bar q} - \bar q \dot{\bar p} )
= \del_s (- r^2 \dot \pht)
\label{partial}
\ee
Thus $\delta S_{kin}$ appearing in \eq{action-for-delta-u}
agrees with the corresponding term in \eq{path-for-giant-arbit-1}
apart from a proportionality constant $A$. 

Let us discuss the constant $A$.  In the last line of \eq{tmp}
$\delta_q(0)$ denotes $\delta(x_1 - x_1)$, similarly $\delta_p(0)$
denotes $\delta(x_2 - x_2)$.  Clearly we need a regularization. It is
natural to choose $\delta_q(0) = \delta_p(0) = a/\sqrt{\hbar}$. We get
$A=1$ if $a^2 = \pi/3$. We do not believe that this regularization has
a particular significance since the agreement at the level of the
equation of motion, between \eq{sho} and \eq{eom-delta-u}, does not
use any such regularization. In other words, the equation of motion
\eq{eom-delta-u}, which can be derived from \eq{sho}, can be used to
fix the relative coefficients between $\delta S_{kin}$ and $\delta
S_{ham}$ in 
\eq{action-for-delta-u}, thus 
determining $A=1$ in \eq{tmp}. Such a method proves the desired result
without the use of a regularization. 

\section{\label{graviton}Gravitons}

So far in this paper we have primarily 
considered collective motions identified as D3
branes. We found that (see \eq{action-arbit}) the $\hbar$ of 
the collective action
naturally corresponds to the D3-brane tension:
\be
\frac1{2\hbar} =  T_3 \omega_3
\label{d3-tension-match}
\ee
This raises a puzzle about other collective motions
such as 
gravitons. Suppose we consider an equation
analogous to \eq{main-eqn}, where the $\delta u$
fluctuation corresponds to a ripple (see
footnote \ref{tiling}) and the brane refers 
now to a fundamental string. Since the left hand
side of \eq{main-eqn} continues to have a prefactor $1/\hbar$
(see, e.g \eq{action-for-delta-u}, 
\eq{delta-s-ham}), while 
the fundamental string tension does not
involve $1/g_s$, we apparently have a puzzle here.

The resolution comes from the fact that $\delta u$
now describes  ``ripples'' which are
fluctuations extending from the original droplet(s) by distances
$O(\sqrt\hbar)$.  Because of this, as we will  show below, 
the collective action
evaluates to $O(g_s)$ which cancels the $1/g_s$, 
reproducing the fundamental string tension so far
as $g_s$-counting is concerned.

\begin{figure}[ht]
\vspace{0.5cm}
\hspace{-0.5cm}
\centerline{
    \epsfxsize=8.5cm
   \epsfysize=8cm
   \epsffile{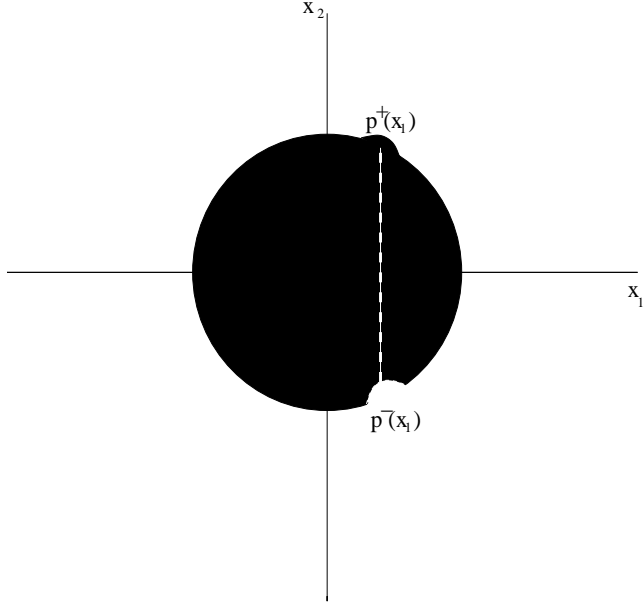}
 }
\caption{\sl The fluctuations $p^+(x_1)$ and $p^-(x_1)$
extend from the original droplet to distances $O(\sqrt\hbar)$.
The Lagrangian for these fluctuations evaluates to $O(g_s)$.
This cancels the prefactor $1/g_s$ sitting outside the collective action
\eq{action-for-delta-u}, consistent with fundamental
string tension which is independent of $g_s$.
}
\label{ripple.fig}
\end{figure}

The simplest parameterization \cite{Polchinski,Sengupta-Wadia}
for the ripples is as in Figure \ref{ripple.fig}.  
For simplicity we have considered the unperturbed 
droplet to correspond to \ads, but similar arguments
can be made with respect to ripples traveling
in other backgrounds.

The precise form of $u(x_1, x_2)$ is
\be
u(x_1, x_2) = \theta([p^+(x_1) - x_2][x_2 - p^-(x_1)])
\ee
where $p^\pm(x_1)$ are to be chosen consistent
with \eq{int-const}.  
The fact that the amplitude of the fluctuations
$\sim O(\sqrt\hbar)$ implies $\delta p^+, \delta p^-
\sim O(\sqrt\hbar)$, where $\delta p^\pm =
p^\pm(x_1) \mp p_0^\pm(x_1)$. $ p_0^\pm(x_1)$ denote
the unperturbed profile. 
Following steps similar
to \cite{Polchinski,Sengupta-Wadia} 
the action $\delta S$ for the fluctuation turns out to be quadratic
in $\delta p^+, \delta p^-$ and hence $\sim O(\hbar)
\propto O(g_s)$. 
Thus, $g_s$ cancels from the left hand side of
\eq{main-eqn} for ripples, consistent with their
interpretation as  fundamental
string modes.

We hope to come back to a quantitative derivation of the action (as well
as path integral) for gravitons from the collective coordinate path
integral \eq{path-for-u}.


\bibliographystyle{utphys} 
\bibliography{myrefs}

\end{document}